\documentclass[conference]{IEEEtran}
\usepackage{cite}
\usepackage{threeparttable} 
\usepackage{array}
\usepackage{subfigure}
\usepackage{tablefootnote}
\usepackage{booktabs}
\usepackage{stfloats}
\usepackage{amsmath,amssymb,amsfonts}
\usepackage{graphicx}
\usepackage{textcomp}
\usepackage{xcolor}
\usepackage{algorithm}  
\usepackage[normalem]{ulem}  
\usepackage{algpseudocode}  
\usepackage{amsmath}  
\usepackage{multirow}

\usepackage{geometry}
\begin{document}
\date{}

\title{\Large\bf
Logic Design of Neural Networks for High-Throughput and Low-Power Applications
\vspace{-0.3cm}}

\author{
\IEEEauthorblockN{Kangwei Xu\textsuperscript{1}, Grace Li Zhang\textsuperscript{2}, Ulf Schlichtmann\textsuperscript{1}, Bing Li\textsuperscript{1}}
\IEEEauthorblockA{\textsuperscript{1}\textit{Chair of Electronic Design Automation, Technical University of Munich (TUM)}, Munich, Germany \\
\textsuperscript{2}\textit{Hardware for Artificial Intelligence Group, Technical University of Darmstadt}, Darmstadt, Germany \\
Email: kangwei.xu@tum.de, grace.zhang@tu-darmstadt.de, ulf.schlichtmann@tum.de, b.li@tum.de}
\vspace{-0.9cm}
}

\maketitle
\thispagestyle{empty}
\vspace{-0.82cm}
{\small\bf Abstract---
 Neural networks (NNs) have been successfully deployed in various fields. In NNs, a large number of multiply-accumulate (MAC) operations need to be performed. Most existing digital hardware platforms rely on parallel MAC units to accelerate these MAC operations. However, under a given area constraint, the number of MAC units in such platforms is limited, so MAC units have to be reused to perform MAC operations in a neural network. Accordingly, the throughput in generating classification results is not high, which prevents the application of traditional hardware platforms in extreme-throughput scenarios. Besides, the power consumption of such platforms is also high, mainly due to data movement. To overcome this challenge, in this paper, we propose to flatten and implement all the operations at neurons, e.g., MAC and ReLU, in a neural network with their corresponding logic circuits. To improve the throughput and reduce the power consumption of such logic designs, the weight values are embedded into the MAC units to simplify the logic, which can reduce the delay of the MAC units and the power consumption incurred by weight movement. The retiming technique is further used to improve the throughput of the logic circuits for neural networks. In addition, we propose a hardware-aware training method to reduce the area of logic designs of neural networks. Experimental results demonstrate that the proposed logic designs can achieve high throughput and low power consumption for several high-throughput applications.}

\section{Introduction}
Neural networks (NNs) have been successfully applied in various fields, e.g., pattern recognition and natural language processing. In NNs, a large number of multiply-accumulate (MAC) operations need to be performed. Traditional digital hardware platforms such as GPU and TPU use parallel MAC units consisting of multipliers and adders to perform such MAC operations. Due to area constraints, the number of MAC units is limited. Accordingly, MAC units on such platforms have to be reused to implement all the MAC operations in a neural network.
Therefore, the throughput of generating classification results on such platforms is not high, which prevents their adoption in extremely high-throughput applications, such as signal compensation in optical fiber communications~\cite{b1}, data collection from physics experiments~\cite{b3} and malicious packet filtering for network detection~\cite{b3}. In addition, large power consumption is another issue using such platforms to accelerate NNs, mainly resulting from data movement, e.g., loading weights from external DRAM to MAC units~\cite{b3.1}. MAC operations also cause a part of the power consumption. Thus it remains challenging to perform high-throughput tasks on those resource-constrained platforms requiring low power consumption. 

Various methods, from the software to the hardware levels, have been proposed to address the throughput and power consumption issues in traditional digital hardware platforms. On the software level, pruning~\cite{b4} and quantization~\cite{b4.1} of NNs have been explored to reduce the number of MAC operations and the complexity of performing MAC operations, respectively. In addition, different dataflows, e.g., weight stationary~\cite{b5}, output stationary~\cite{b6}, and row stationary~\cite{b7}, have been proposed to reduce data movement and power consumption. On the hardware level, traditional digital hardware platforms using parallel MAC units are modified, e.g., by inserting multiplexers~\cite{b8} to improve the throughput, and power/clock gating~\cite{b9}~\cite{b10} to reduce power consumption.  

Another perspective to improve throughput and reduce power consumption in accelerating a neural network is to convert it into a logic circuit or a look-up table (LUT)-based design, where weights are embedded. For example, LogicNets~\cite{b3} quantizes the inputs and outputs of neurons with low bit widths and implements such neurons with LUTs, which are subsequently deployed on FPGAs. NullaNets~\cite{b10.1} train a neural network to produce binary activations and treat the operations at a neuron as multi-input multi-output Boolean functions. Such Boolean functions of neurons are further considered as truth tables and synthesized with logic synthesis tools. 

Although converting a neural network into a logic/LUT design is promising to improve throughput and reduce power consumption in accelerating this neural network, the existing methods either incur a large number of LUTs for a high inference accuracy or cannot guarantee the feasibility of logic synthesis of truth tables of neurons due to their complexity. To address this challenge, in this paper, we propose to directly flatten and implement all the operations in a neural network with their corresponding logic circuits and embed weights into such circuits. The key contributions of this paper are summarized as follows. 

\begin{itemize}
\item %
All operations including MAC and activation functions in a neural network are flattened and implemented with logic circuits. Flip-flops are inserted at the end of neurons in each layer to improve the throughput of such circuits. 

\item 
In implementing MAC operations with logic circuits, predetermined weights after training are used to simplify the logic of MAC units to reduce their delay, power and area. Since weights are not required to be moved from external DRAM, power consumption can be reduced significantly.  

\item After the MAC units are simplified with weights in a layer, some logic is shared between the layers. Therefore, the whole neural network circuit is further simplified with EDA tools to reduce area and delay.

\item 
Different weight values affect the logic complexity of the resulting simplified MAC units. Accordingly, the traditional training is adjusted to select those weight values leading to smaller circuit sizes.

\item We present comprehensive evaluations on three high-throughput tasks, demonstrating that the proposed method can achieve high throughput and low power consumption. Furthermore, the proposed method outperforms the state-of-art approaches by achieving an average 2.19\% increase in inference accuracy and an average 13.26\% increase in throughput while reducing area overheads by an average of 38.76\%.
\end{itemize}

The rest of this paper is organized as follows. Section~\ref{sec:second} gives the background and motivation of this work. Section~\ref{sec:third} explains the details of the proposed method. The experimental results are given in Section~\ref{sec:fourth}. Section~\ref{sec:fifth} concludes the paper.

\begin{figure}[]
\centering
	\includegraphics[width=0.9\linewidth]{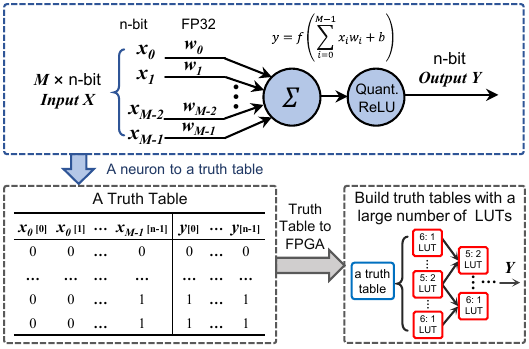}
	\vspace{-0.2cm}
	\caption{~The concept of implementing a neuron with LUTs in LogicNets~\cite{b3}.}
	\label{fig:fpga}
	\vspace{-0.42cm}
\end{figure}

\section{Background and Motivation}\label{sec:second}
 
Neural networks have multiple layers of neurons. Synapses connect neurons with individual weights. The weights of a layer form a weight matrix. Computing a layer in an NN requires the multiplication of input data and the weight matrix, incurring a large number of MAC operations. The results of MAC operations are further processed by activation functions, e.g., ReLU. 

Traditional digital hardware platforms such as GPU and TPU adopt parallel MAC units to accelerate such MAC operations. Due to the area constraint of such platforms, the number of MAC units is limited, so those MAC units have to be reused to perform all the MAC operations. Therefore, the throughput of generating classification results is low and not well suited for extreme-throughput applications. In addition, such platforms suffer from large power consumption due to data movement and MAC computation, limiting their application in resource-constrained platforms, e.g., edge devices. 

Various techniques have been proposed to address throughput and power consumption issues in accelerating NNs. One of the promising techniques is to convert NNs into LUT-based designs and logic circuits. We will explain their basic concepts as follows. 

\textit{1) LogicNets:} \cite{b3} proposes to map quantized neurons in a neural network to LUTs and implement such LUTs on FPGAs. The concept of LogicNets is shown in Fig.~\ref{fig:fpga}, where the $M$ inputs and one output of a neuron are quantized to $n$ bits. To convert this neuron into a LUT, the function of neurons can be expressed by a truth table that enumerates all the combinations of input bits (i.e., the fan-in of this neuron) with their corresponding outputs. For an entry in the truth table, the output can be evaluated according to the input bits and predetermined weights. The weights themselves do not appear in the truth table directly. Once the truth table is established, it can be mapped to LUTs and implemented on FPGAs.

Since neurons in NNs may have many inputs and their quantization bits should be large enough to maintain a high inference accuracy. As the truth table becomes large, the above mapping may lead to a large number of LUTs. This is because the hardware cost of implementing truth tables of neurons with LUTs grows exponentially with neuron fan-in. For example, implementing a 32:1 truth table requires about a hundred million 6:1 LUTs, which is much larger than even the largest FPGAs available today and makes it impractical for direct mapping to LUTs~\cite{b3}. Although extreme pruning and quantization can limit the neuron fan-in, this degrades the inference accuracy.

\textit{2)  NullaNets:} 
\cite{b10.1} converts the truth table of a neuron into a logic circuit with logic synthesis. To reduce the complexity in logic synthesis, they only evaluate the output values for input combinations in the training dataset, and the output values for the remaining input combinations are set as ``don't care". The resulting synthesized logic circuit may not generate the correct output for those input data in the test dataset, degrading the inference accuracy. Besides, even though only partial input combinations are considered in logic synthesis, the logic complexity might still be high, preventing the feasibility of logic synthesis of the truth table of a neuron. Contrary to the previous work in converting NNs into LUTs and logic circuits, the proposed method directly flattens all the operations in a neural network into their corresponding logic circuits with weights embedded in such circuits. 

\section{Logic Design of Neural Networks}\label{sec:third}
In this section, we first introduce the logic implementations of neural networks. Retiming is then used to improve the throughput of the logic designs of NNs. Afterwards, a hardware-aware training technique is proposed to reduce the area overheads of such logic designs. 

\begin{figure}[]
\centering	\includegraphics[width=1.01\linewidth]{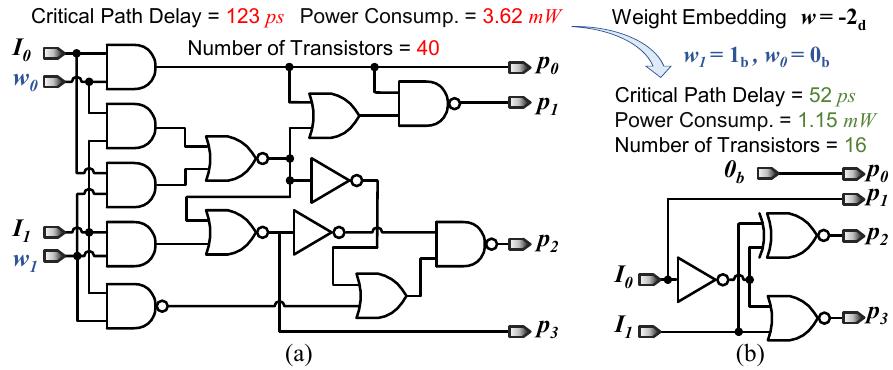}
	\vspace{-0.7cm}
	\caption{~Logic circuit of 2-bit signed multiplier. (a) The original circuit; (b) The logic circuit simplified with a fixed quantization weight (decimal: -2, binary: 10).}
	\label{fig:2bitmultiplier}
	\vspace{-0.6cm}
\end{figure}

\subsection{Logic Implementations of NNs with Fixed Weights}

To implement all the operations in NNs with logic circuits, we first use quantization-aware training to train a neural network while maintaining the inference accuracy. 8-bit quantization of weights and input activations is used during this process~\cite{b11}. In addition, unstructured pruning~\cite{b12} is further used to remove unimportant weights and reduce computational costs. Afterwards, the neural network is fine-tuned to improve the inference accuracy.

After training, the MAC operations in a neural network can be directly implemented with MAC units and the logic realizing the activation functions. The fixed weights after training are used to simplify the MAC operations at neurons. The simplified MAC units are appended with the logic circuit implementing the activation function at a neuron. All the neurons in this network are processed similarly. The resulting logic circuits are concatenated to generate the complete logic of the neural network. Then, flip-flops are inserted at the end of each layer to synchronize data propagation, after which logic redundancy within and across layers is removed by EDA tools. In the following paragraphs, we will introduce each part of the logic design of a neural network.

\textit{1) Multiplier:} To reduce the delay, power and area of multipliers, the fixed weights after training are used to simplify the logic of the multipliers. Fig.~\ref{fig:2bitmultiplier} illustrates the logic simplification of a 2-bit signed multiplier with the fixed quantized weight -2$_{d}$ or 10$_{b}$. After this simplification, the delay of the multiplier circuit is reduced by 57.72\%, the power consumption is reduced by 68.23\%, and the number of transistors is reduced by 60\%. All the multipliers at a neuron are processed this way. 

\begin{figure}[]
\centering
	\includegraphics[width=1.01\linewidth]{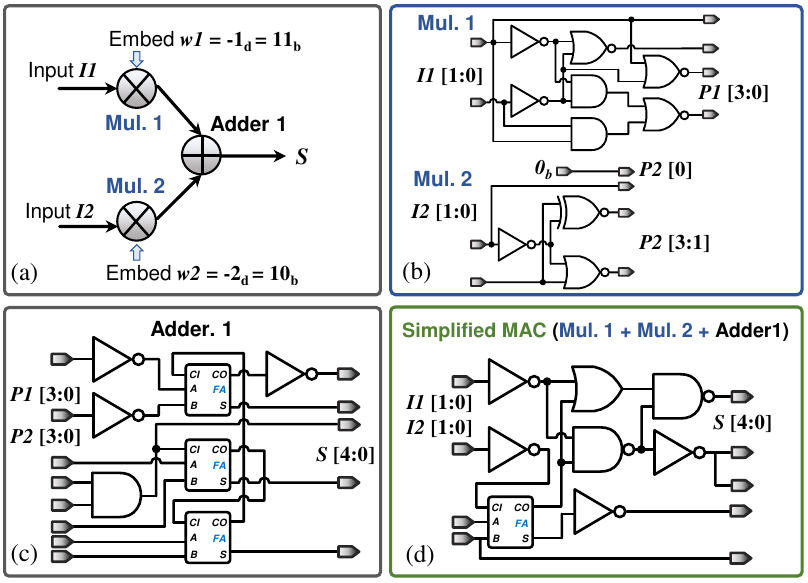}
	\vspace{-0.55cm}
	\caption{~(a) MAC operations at a neuron; (b) 2-bit signed multipliers simplified with the fixed quantized weights; (c) 4-bit signed adder circuit before logic simplification, where FA is a 1-bit full adder; (d) Circuit of the simplified MAC unit.}
	\label{fig:2bitmac}
	\vspace{-0.3cm}
\end{figure}

\textit{2) Adder:} 
The simplified multipliers at a neuron will be appended with an adder realizing the addition operation. Due to the weight embedded in multipliers, the resulting logic circuit of the MAC operation can be further simplified. Fig.~\ref{fig:2bitmac} illustrates the comparison of the MAC unit circuit before and after weight embedding, where the multipliers and the adder are 2-bit and 4-bit, respectively. With weight embedding, the delay of the MAC unit is reduced by 70.07\%, the power consumption is reduced by 71.64\%, and the number of transistors is reduced by 65\%.

In this work, we use an 8-bit neural network to maintain the inference accuracy of the hardware implementation. However, the quantization bit of the adder of a MAC unit is larger than 8-bit and increases with the increasing number of inputs at a neuron. In general, the quantization bit of the MAC operation result should be reduced to 8-bit before this result enters the subsequent layer~\cite{b11}. Two techniques are proposed to maintain the inference accuracy during this process.

\begin{figure}[]
\centering	\includegraphics[width=1\linewidth]{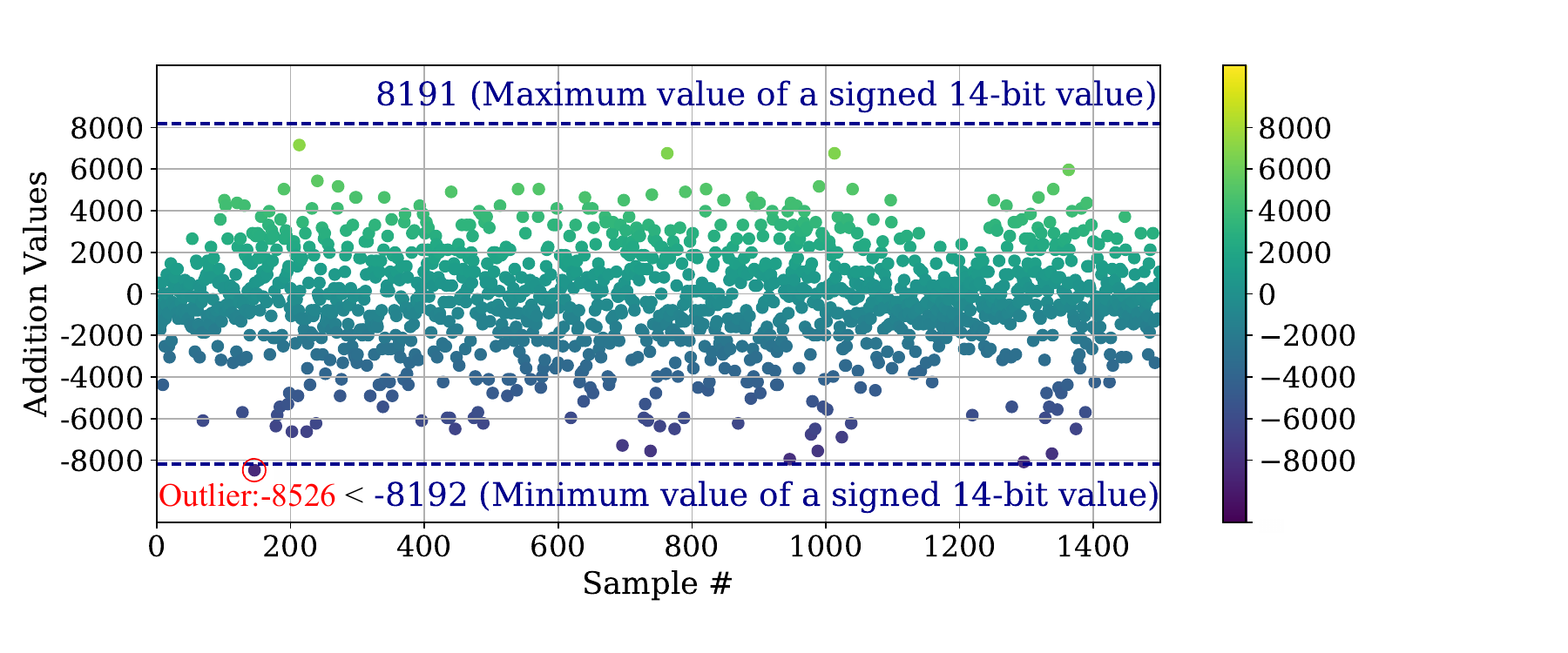}
	\vspace{-0.6cm}
	\caption{~Distribution of 1500 samples of addition values from the adder output in a neuron for the OFC task. This example illustrates that after removing the outlier -8256${_d}$ (15-bit), the output bit width of the adder needs only 14-bit.}
	\label{fig:adderbitwidth}
	\vspace{-0.45cm}
\end{figure}

First, all the addition results at neurons in NNs are profiled to examine the actual bit width to represent such addition results. Fig.~\ref{fig:adderbitwidth} presents an example of the distribution of addition results according to the training dataset. By removing the outliers, the maximum values of the remaining results are used to define the output bit width of the adder. Second, a requantizer is used to convert the high-bit value to 8-bit by multiplying it with a scaling factor~\cite{b11}. By embedding the scaling factor into the requantizer, the delay, power and area of the requantizer circuit can be reduced. 

\textit{3) The Circuit for Implementing Activation Function:} An activation function outputs the corresponding activation state by thresholding the input to determine whether the neuron should be activated or not activated. The activation function ReLU generates 0 if the input is smaller than 0 and remains the original value if it is larger than 0. The circuit for performing ReLU is generated by describing its function with hardware description languages such as Verilog and synthesizing it by the EDA tool. 

The logic circuit of each neuron is generated with the techniques described above. The resulting circuits for all neurons are concatenated together to produce the complete logic design of the neural network. The logic circuits of neurons might share some common logic since they result from the simplification of a MAC unit. To remove the logic redundancy, the complete circuit of a neural network will be optimized by EDA tools.  

\begin{figure}[]
\vspace{-0.22cm}
\centering	\includegraphics[width=0.945\linewidth]{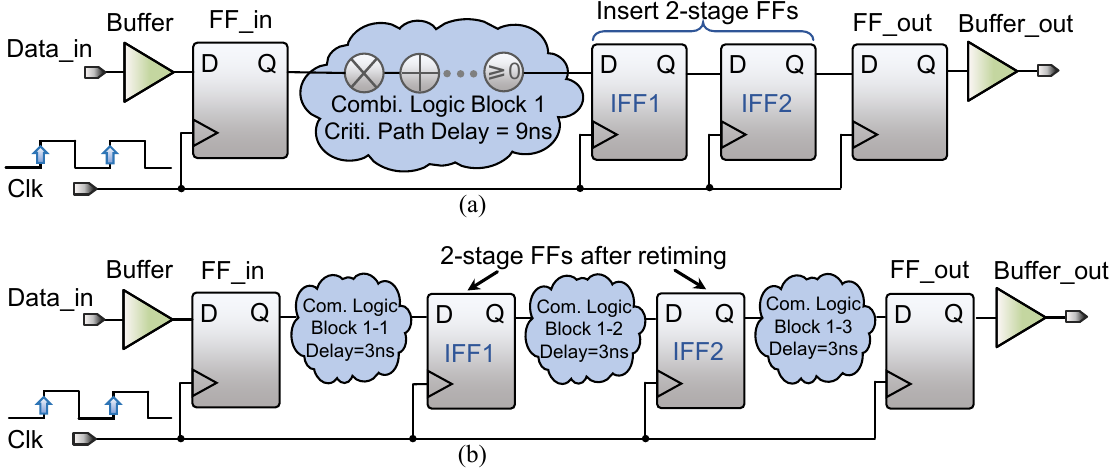}
	\vspace{-0.3cm}
	\caption{~(a) Insertion of two-stage flip-flops after the original ReLU circuit; (b) Circuit after retiming.}
	\label{fig:retiming}
	\vspace{-0.62cm}
\end{figure}

\subsection{Retiming with Cascaded Flip-Flops}
To synchronize data propagation in each layer, 
flip-flops are inserted at the end of the logic circuit implementing the activation function. To further improve the throughput of the logic circuit of a neural network, cascaded flip-flops are inserted into the original circuits, and retiming technique is then deployed to reduce the maximum delay between flip-flop stages. As a retiming example at a neuron shown in Fig.~\ref{fig:retiming}(a), the combinational logic block 1 has a critical path delay of 9, limiting the whole circuit's performance. Assume the clock-to-q delay and the setup time of a flip-flop are 3 and 1, respectively. The minimum clock period of this circuit is equal to 13. In this example, two-stage flip-flops (IFF1 and IFF2) are inserted after the circuit implementation of ReLU. Afterwards, retiming is realized by EDA tools automatically. As shown in Fig.~\ref{fig:retiming}(b), with the retiming technique, the performance can be optimized by reallocating IFF1 and IFF2 in the combinational logic block 1, resulting in a minimum clock period equal to 7, which reduces the clock period by 46.15\%.

\subsection{Hardware-Aware Training}
\begin{figure}[]
\vspace{-0.22cm}
\centering	\includegraphics[width=0.94\linewidth]{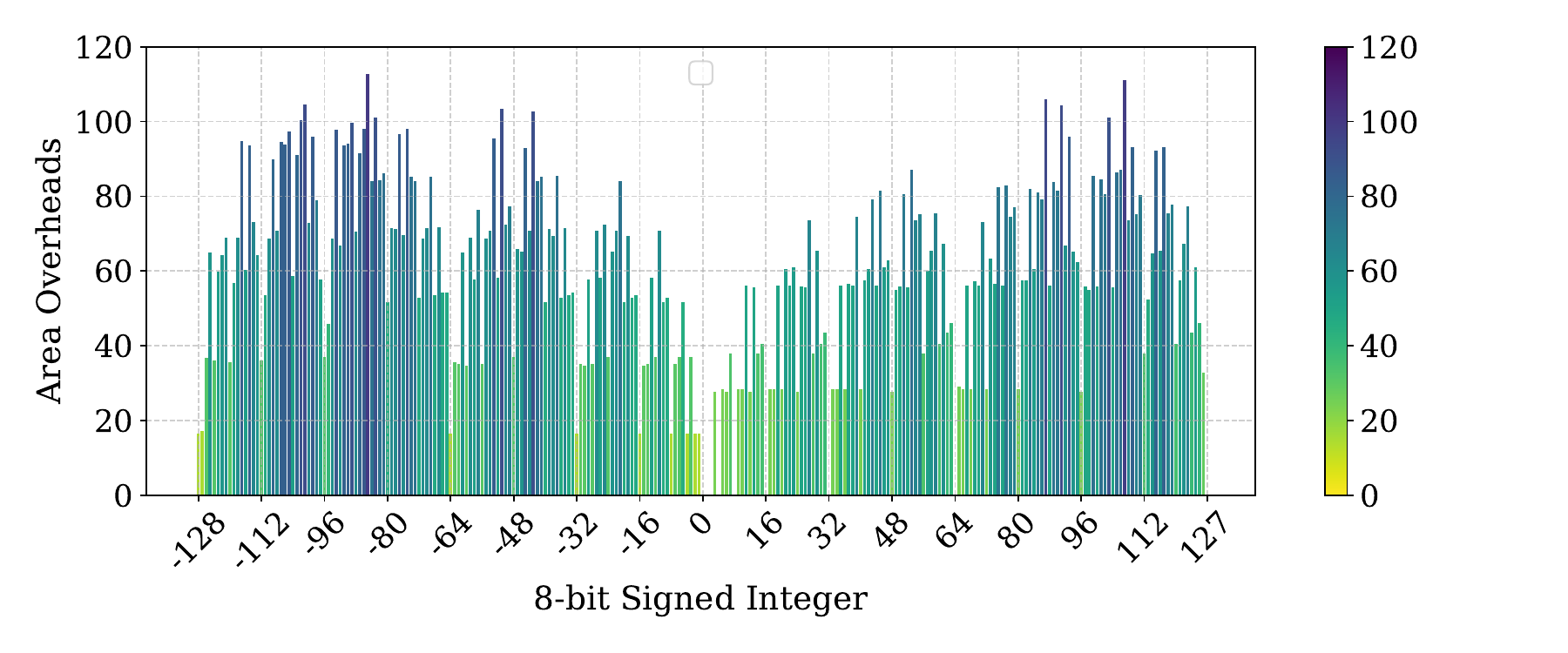}
	\vspace{-0.28cm}
	\caption{Area of multipliers simplified with 8-bit quantized weights.}
	\label{fig:mula}
	\vspace{-0.42cm}
\end{figure}

Different weight values affect the logic complexity of the resulting simplified multiplier. For example, as shown in Fig~\ref{fig:mula}, the quantized 8-bit weight ‘107’ corresponds to a larger multiplier area of 111 units, while the quantized 8-bit weight ‘-16’ leads to an area overhead of only 16 units. To take advantage of this property, we propose to train the neural network with selected weights that lead to a smaller multiplier area. The weight selection and the modified training are explained as follows. 

\textit{1) Weight selection:} We first rank the weight values according to the area of the resulting simplified multipliers. Then, we select the top $n$ weights that lead to the smallest multiplier area. In the experiments, $n$ was set to 40. The neural network is trained only with such weight values, and the validation accuracy is verified. If the validation accuracy is much lower than that of the original training, more weight values, e.g., 50, are selected and used to train the neural network. In each iteration, 10 more weight values leading to the small area will be added to the previously selected weight value set and used to train the network. The iteration continues until the validation accuracy recovers almost to that of the original training.

\textit{2) Training:} During training, the weights are forced to take the selected values in the forward propagation. In the backward propagation, the straight-through estimator is applied to skip the selection operation. In each epoch, all weights in the model are traversed and replaced with the closest value from the selected weight set. The training continues until the loss function converges or a given number of epochs is finished.

\begin{table*}[!t]
\vspace{-0.0cm}
  \centering
  \caption{The performance result with the proposed method. }
  \begin{threeparttable}
    \fontsize{9pt}{9pt}\selectfont
    \resizebox{\textwidth}{!}{%
      \begin{tabular}{
      l
      l
      p{1.1cm}
      p{1.1cm}
      p{1.1cm}
      p{0.65cm}
      p{0.65cm}
      p{1.1cm}
      p{0.65cm}
      p{0.65cm}
      p{1cm}
      p{0.7cm}
      p{0.7cm}}
      \toprule[1.2pt]
      \multicolumn{1}{l}{\multirow{2.5}{*}{\textbf{Network}}} & \multicolumn{1}{l}{\multirow{2.5}{*}{\textbf{Neurons / Layer}}}
      & \multicolumn{3}{c}{\textbf{BER(OFC) or Acc.(JSC/NID)\tnote{1}}} & \multicolumn{3}{c}{\textbf{Max. Frequency (MHz)}} & \multicolumn{3}{c}{\textbf{Power Consum. (mW)}} & \multicolumn{2}{c}{\centering\textbf{No. of Weight}} \\
      \cmidrule(lr){3-5} \cmidrule(lr){6-8} \cmidrule(lr){9-11} \cmidrule(l){12-13}
      & & \textbf{Float} & \textbf{Pruned} & \textbf{Prop.} & \textbf{Base.} & \textbf{Retim.} & \textbf{Improv.} & \textbf{Base.} & \textbf{Prop.} & \textbf{Drop} & \textbf{Base.} & \textbf{Prop.} \\
      \midrule \vspace{2pt}
      OFC-A & \multicolumn{1}{l}{21, 40, 25} & 4.34${e^{-4}}$ & 5.43${e^{-4}}$ & 5.87${e^{-4}}$ & 338 & 568 & 68.05\% & 543 & 88 & 83.79\% & 256 & 70 \\ \vspace{2pt}
      OFC-B & \multicolumn{1}{l}{21, 50, 25} & 3.26${e^{-4}}$ & 3.91${e^{-4}}$ & 4.13${e^{-4}}$ & 322 & 510 & 58.39\% & 622 & 102 & 83.60\% & 256 & 80 \\ 
      OFC-C & \multicolumn{1}{l}{21, 50, 50} & 2.39${e^{-4}}$ & 2.61${e^{-4}}$ & 3.26${e^{-4}}$ & 306 & 463 & 51.31\% & 986 & 122 & 87.63\% & 256 & 80 \\
      \midrule \vspace{2pt}
      JSC-A & \multicolumn{1}{l}{16, 64, 16, 16, 8} & 73.94\% & 73.69\% & 73.65\% & 364 & 564 & 54.95\% & 763 & 181 & 76.28\% & 256 & 80 \\ \vspace{2pt}
      JSC-B & \multicolumn{1}{l}{16, 64, 32, 32, 32} & 74.42\% & 74.20\% & 74.11\% & 336 & 442 & 31.55\% & 1270 & 299 & 76.46\% & 256 & 100 \\ 
      JSC-C & \multicolumn{1}{l}{16, 64, 48, 48, 32} & 74.77\% & 74.50\% & 74.44\% & 305 & 422 & 38.36\% & 1783 & 432 & 75.77\% & 256 & 110 \\ 
      \midrule \vspace{2pt}
      NID-A & \multicolumn{1}{l}{593, 20} & 90.89\% & 90.63\% & 90.34\% & 458 & 538 & 17.47\% & 349 & 120 & 65.62\% & 256 & 100 \\ \vspace{2pt}
      NID-B & \multicolumn{1}{l}{593, 20, 20} & 91.92\% & 91.85\% & 91.69\% & 360 & 495 & 37.50\% & 412 & 139 & 66.26\% & 256 & 100 \\ 
      NID-C & \multicolumn{1}{l}{593, 25, 25} & 92.07\% & 91.89\% & 91.76\% & 352 & 481 & 36.65\% & 496 & 164 & 66.94\% & 256 & 120 \\
      \bottomrule[1.2pt]
      \end{tabular}%
    }
    \begin{tablenotes}
      \item[1] \scriptsize In general, the Bit Error Rate (BER) indicates the inference performance of the OFC task, while the accuracy (Acc.) indicates the inference performance of the JSC and NID tasks.
    \end{tablenotes}
    \label{tab:addlabel}%
  \end{threeparttable}
\vspace{-0.1cm}
\end{table*}

\section{Experimental Results}\label{sec:fourth}
In this section, we demonstrate the results of the proposed method in terms of accuracy, throughput, power consumption and area overheads in 3 different extreme-throughput applications:

\textit{1) Optical Fiber Communications (OFC):} Transmission of optical signals in fibers suffers from chromatic dispersion (CD). Neural networks are used as nonlinear equalizers to compensate CD, where the input is the optical signal affected by CD, and the output is the compensated optical signal. We use the formulation from~\cite{b1} for OFC as a 21-input 1-output prediction task.

\textit{2) Jet Substructure Classification (JSC):} In large-scale physics experiments, terabytes of instrumentation data are generated every second. Neural networks with 16 inputs and 5 outputs are employed to filter out the most interesting results~\cite{b2}.

\textit{3) Network Intrusion Detection (NID):} The neural networks can identify malicious network packets to strengthen network security. For this task, we use the UNSW-NB15 dataset~\cite{b3}, which consists of the packets labeled as either bad (0) or normal (1) with a total of 593 input features.

During training, the Adam optimizer was used and the step decay learning rate schedule was set starting from 0.001. The mini-batch size was set to 1024. The quantization-aware training stops after 300 epochs and the hardware-aware training stops after 100 epochs. Each experiment for a particular task was repeated 10 times and the average is reported. The neural network training was implemented with Pytorch on Nvidia Quadro RTX 6000 GPUs. The logic circuits in implementing NNs were synthesized
with Synopsys Design Compiler using the Nangate 45nm open-cell library. The correctness of the circuits is verified by performing simulation and ensuring the same results as the original PyTorch network are returned.

\begin{figure*}[!t]
\vspace{-0.1cm}
    \centering
    \begin{minipage}{0.725\textwidth}
        \centering
        \includegraphics[width=\linewidth]{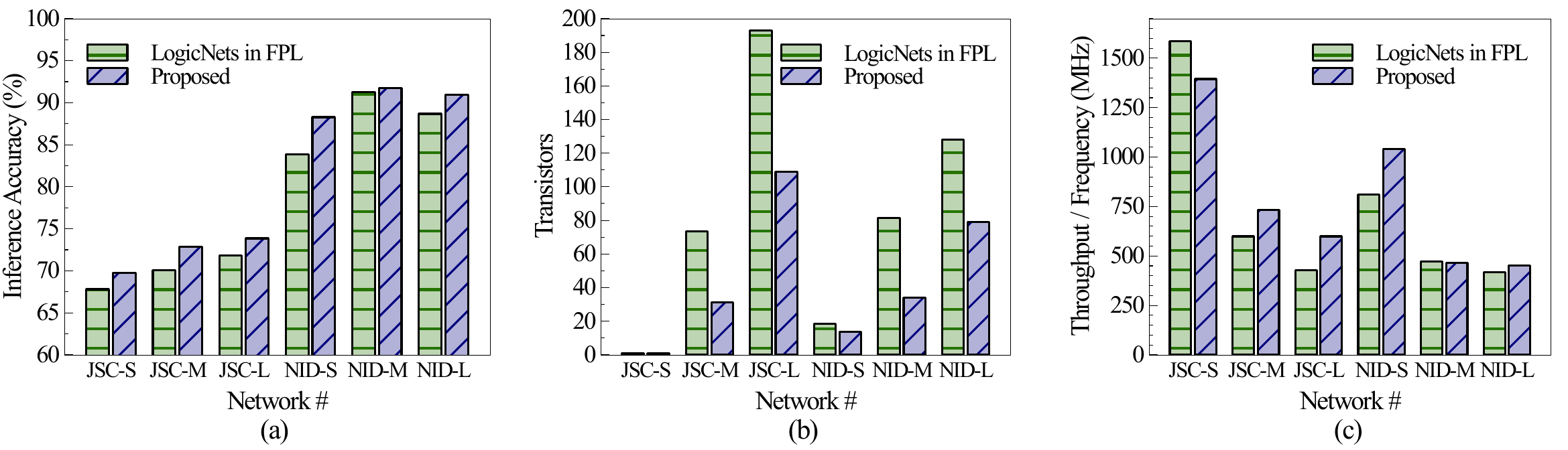}
        \vspace{-0.72cm}
        \caption{(a) Comparison of inference accuracy with LogicNets; (b) Comparison in the number of transistors in the hardware implementation with LogicNets; (c) Comparison of maximum frequency with LogicNets.}
        \label{fig:fpl}
        \vspace{-0.5cm}
    \end{minipage}%
    \hspace*{0.2cm}
    \begin{minipage}{0.25\textwidth}
        \centering
        \includegraphics[width=\linewidth]{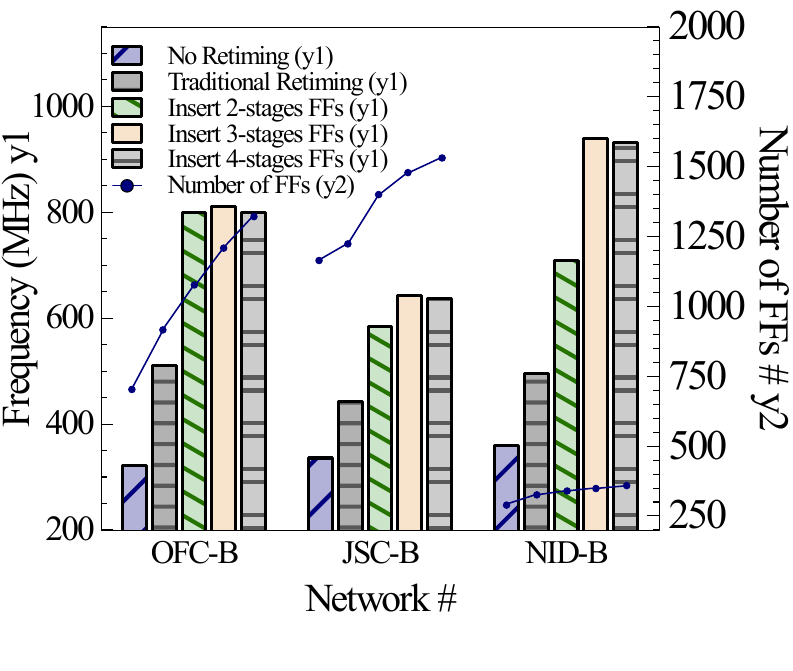}
        \vspace{-0.77cm}
        \caption{Comparison of max frequency before \& after FF insertion and retiming.}
        \label{fig:timing1}
        \vspace{-0.5cm}
    \end{minipage}
\end{figure*}

Table I demonstrates the performance results. The first and second columns show the names of the neural networks and their structures. ``A", ``B" and ``C" represent different versions of neural networks. The third, fourth and fifth columns represent the inference performance of the original floating-point NNs, the traditional quantized and pruned NNs, and our proposed NNs, respectively. According to these columns, the proposed hardware-aware training method can maintain a low bit error rate (BER) in the application of OFC and high inference accuracy in the applications of JSC and NID. Compared with the baseline, where all the operations in a neural network are flattened without embedding fixed weights, the proposed method can achieve higher throughput and lower power consumption, as shown in the eighth and eleventh columns. The last two columns show the numbers of selected weight values in the hardware-aware training. Compared with the original 8-bit of 256 weight values used for training, we train the NNs with only a small number of weight values (e.g., 70 in OFC-A) to reduce the size of their logic implementations. 

To demonstrate the advantages of the proposed method, we compared the proposed method with LogicNets~\cite{b3} in terms of inference accuracy, the number of transistors and maximum frequency. In this comparison, the network size, the pruning ratio, and the quantization bits of the activations were set as the same with LogicNets. The number of LUTs in LogicNets is equivalent to the number of transistors for area comparison. According to the results illustrated in Fig.~\ref{fig:fpl}, the proposed method achieves higher accuracy than LogicNets on all tasks with a much smaller number of transistors. In addition, in most cases, the maximum frequency with the proposed method is higher than that of LogicNets, even though LogicNets uses 16nm technology while the proposed method uses 45nm technology.

To balance the maximum frequency and the number of inserted flip-flops, we iteratively inserted more flip-flops after implementing the circuit ReLU and used retiming to improve the clock frequency. The results are illustrated in Fig.~\ref{fig:timing1}, where the histogram denotes the maximum frequency (MHz), and the line represents the number of flip-flops after retiming. According to this figure, the maximum frequency can be improved significantly by inserting more flip-flops initially and later remaining stable even with more flip-flops inserted into the logic circuit. 

\begin{figure}[!t]
\vspace{-0.3cm}
\centering
	\includegraphics[width=1.03\linewidth]{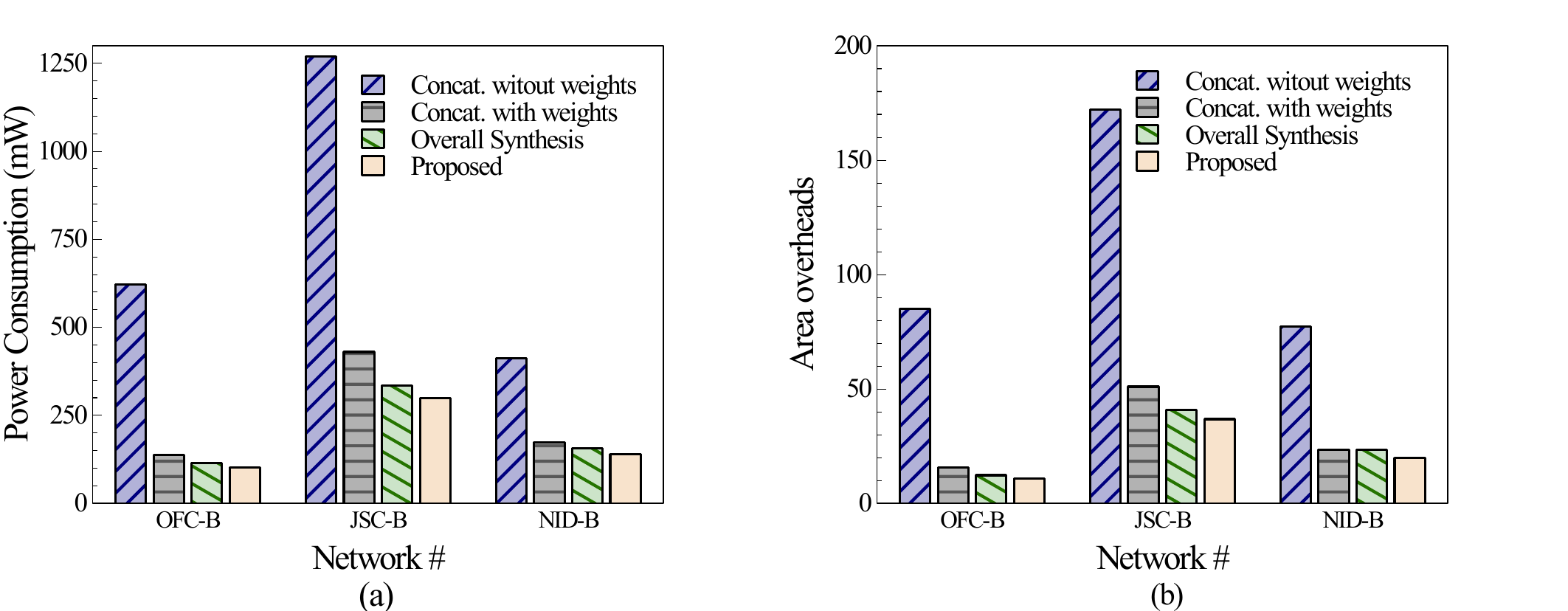}
	\vspace{-0.8cm}
	\caption{~Comparison of circuit performance. (a) Power consumption; (b) Area overheads.}
	\label{fig:cnw1}
	\vspace{-0.65cm}
\end{figure}

To demonstrate the reduction in power and area with weight embedding and hardware-aware training, we compared the power consumption and area overheads before and after using such techniques. As shown in Fig.~\ref{fig:cnw1} (a), for the OFC task, when weights are embedded in the circuit, the power consumption of the synthesized circuit is 137mw. After considering the simplification between different logic, the power consumption is reduced to 115mW. With the hardware-aware training, the power consumption is further reduced to 102mW. According to this figure, weight embedding and hardware-aware training can significantly reduce power consumption and the corresponding area overheads.

\begin{figure}[!t]
\vspace{-0.3cm}
\centering
	\includegraphics[width=1.02\linewidth]{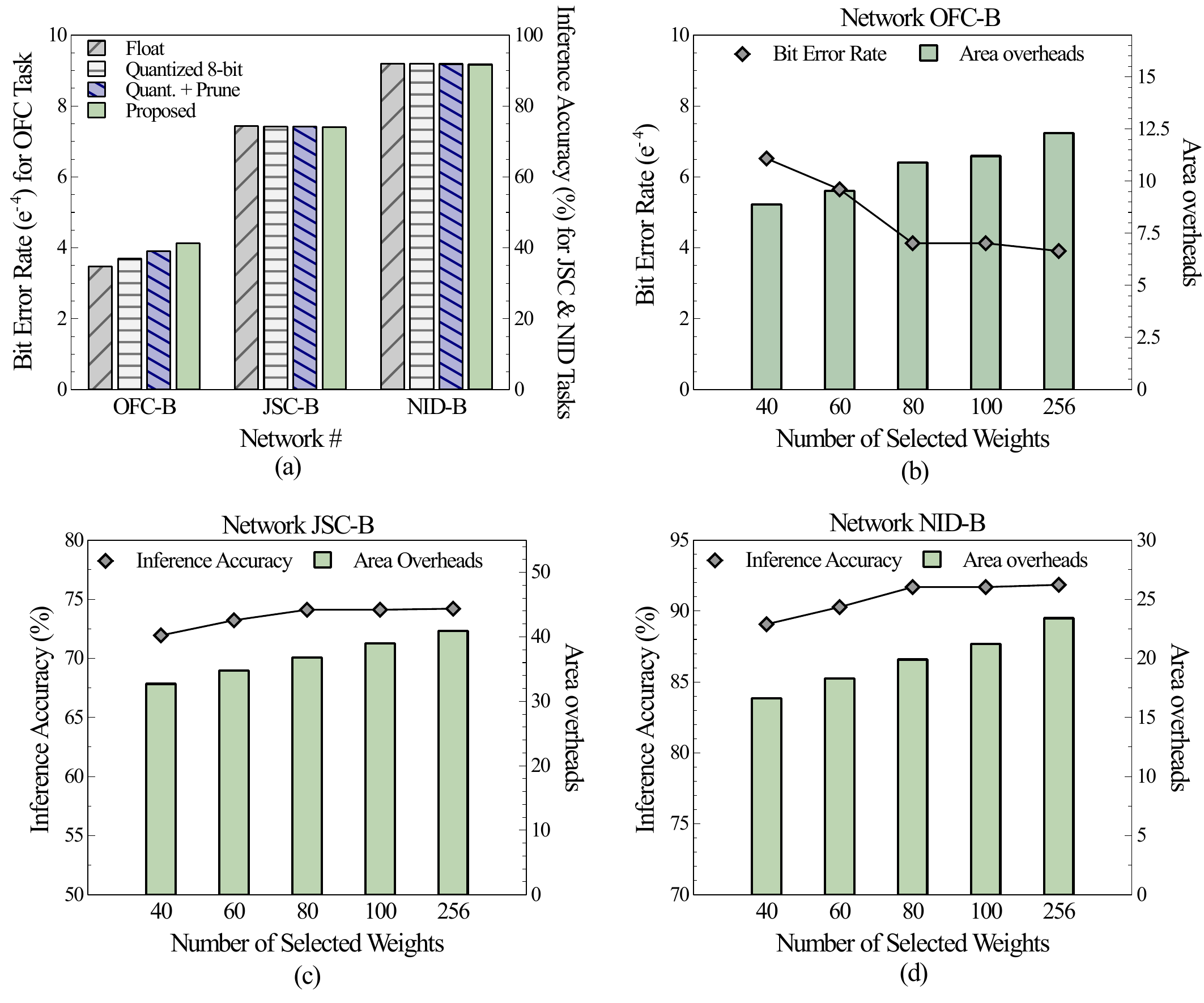}
		\vspace{-0.8cm}
	\caption{~(a) Accuracy Comparison with traditional quantization and pruning; (b)(c)(d) Tradeoff between accuracy, area overheads and the number of selected weights.}
	\label{fig:sw}
	\vspace{-0.75cm}
\end{figure}

In the proposed hardware-aware training, a certain number of weights resulting in a small multiplier area were selected to train the network. The inference accuracy with the proposed method can still be maintained, as illustrated in Fig.~\ref{fig:sw} (a).  
To demonstrate the tradeoff between the number of selected weights and inference performance, we selected different numbers of weights and used them to train the network. The results are illustrated in Fig.~\ref{fig:sw} (b-d). According to these figures, with the number of selected weights reduced, the BER in OFC increases and the inference accuracy in JSC and NID decreases. Besides, the area overheads reduce with decreasing number of selected weights. 

\section{Conclusion}\label{sec:fifth}
In this paper, we have proposed an efficient logic design of neural networks for extremely high-throughput applications. Instead of relying on parallel MAC units to accelerate these MAC operations, all the operations in NNs are implemented through logic circuits with embedded weights. Weight embedding not only reduces the delay of the logic circuit but also reduces power consumption. The retiming technique is used to further improve the circuit throughput. A hardware-aware training method is proposed to reduce the area of logic designs of NNs. Experimental results on three tasks demonstrate the proposed logic design can achieve a high throughput and low power consumption.

\section*{Acknowledgement}
This work is funded by the Deutsche Forschungsgemeinschaft \textit{(DFG, German Research Foundation)} – Project-ID 504518248 and supported by TUM International Graduate School of Science and Engineering \textit{(IGSSE)}.

\end{document}